\documentclass[journal]{IEEEtran}
\usepackage{cite}

\ifCLASSINFOpdf
   \usepackage[pdftex]{graphicx}
\else
\fi

\usepackage{amsmath,amssymb,amsfonts,bm}

\ifCLASSOPTIONcompsoc
  \usepackage[caption=false,font=normalsize,labelfont=sf,textfont=sf]{subfig}
\else
  \usepackage[caption=false,font=footnotesize]{subfig}
\fi

\usepackage{multirow}
\usepackage{booktabs}
\usepackage{array}
\usepackage{threeparttable}

\usepackage[]{algorithm2e}
\makeatletter
\renewcommand{\@algocf@capt@plain}{above}
\makeatother
\usepackage{xcolor}

\begin{document}
%
\title{Deep Learning Based Defect Detection for Solder Joints on Industrial X-Ray Circuit Board Images}
\author{
\IEEEauthorblockN{Qianru Zhang\IEEEauthorrefmark{1}, Meng Zhang\IEEEauthorrefmark{1}, Chinthaka Gamanayake\IEEEauthorrefmark{2}, Chau Yuen\IEEEauthorrefmark{2}, Zehao Geng\IEEEauthorrefmark{3}, Hirunima Jayasekara\IEEEauthorrefmark{2}, Xuewen Zhang\IEEEauthorrefmark{3}, Chia-wei Woo\IEEEauthorrefmark{4}, Jenny Low\IEEEauthorrefmark{4}, Xiang Liu\IEEEauthorrefmark{3}\\}
\IEEEauthorblockA{\IEEEauthorrefmark{1}National ASIC Research Center, Southeast University, Nanjing, China\\} 
\IEEEauthorblockA{\IEEEauthorrefmark{2}Singapore University of Technology and Design, Singapore\\}
\IEEEauthorblockA{\IEEEauthorrefmark{3}Peking University, China\\}
\IEEEauthorblockA{\IEEEauthorrefmark{4}Keysight Company, Singapore\\}}
\maketitle

\begin{abstract}
Quality control is of vital importance during electronics production. As the methods of producing electronic circuits improve, there is an increasing chance of solder defects during assembling the printed circuit board (PCB). Many technologies have been incorporated for inspecting failed soldering, such as X-ray imaging, optical imaging, and thermal imaging. With some advanced algorithms, the new technologies are expected to control the production quality based on the digital images. However, current algorithms sometimes are not accurate enough to meet the quality control. Specialists are needed to do a follow-up checking. 
For automated X-ray inspection, joint of interest on the X-ray image is located by region of interest (ROI) and inspected by some algorithms. Some incorrect ROIs deteriorate the inspection algorithm.

The high dimension of X-ray images and the varying sizes of image dimensions also challenge the inspection algorithms. 
On the other hand, recent advances on deep learning shed light on image-based tasks and are competitive to human levels. 
 
In this paper, deep learning is incorporated in X-ray imaging based quality control during PCB quality inspection. 
Two artificial intelligence (AI) based models are proposed and compared for joint defect detection. 

The noised ROI problem and the varying sizes of imaging dimension problem are addressed.
The efficacy of the proposed methods are verified through experimenting on a real-world 3D X-ray dataset. 
By incorporating the proposed methods, specialist inspection workload is largely saved.

\end{abstract}


%
\IEEEpeerreviewmaketitle

\section{Introduction}
%
%
%
%
\IEEEPARstart{A}{s} the method for producing electronic circuits improves, there is an increasing speed of producing electronics products such as cell phones and laptops. 

At the same time, there is a growing risk of defects during printed circuit board (PCB) assembling because of the component size decreasing and the component density increasing.
Therefore, an efficient and accurate quality control system is essential. 

    Some inspection is done by a group of specialists who are trained to manually inspect all sorts of defects. However, different people have different subjective determinations and their criteria may change as people get fatigue.
    There are many solder joint defects in PCB assembling, among which structural defect is one of the dominant defect types on an assembled circuit board \cite{leinbach2001and}. It includes defects of insufficient solder, voiding, shorts, and so on. With the help of nondestructive methods such as X-ray imaging, optical imaging, and thermal imaging, structural defect can be inspected by the machines automatically. 

    Research on joint defect detection using thermal and optical imaging inspection has gained a great of interest due to the importance of quality control during electronics production that requires efficiency and accuracy.  Lu \textit{et al.}~\cite{lu2018detection}~proposed to use thermal images combined with machine learning method for micro solder balls detection. Specifically, they applied K-means algorithm to the features extracted from the reconstructed thermal images including the area of the solder balls, the variance of the hot spot and the probability of a high temperature, and the defects are the ones that deviate from the clusters. Huang \textit{et al.}~\cite{huang2019developing}~proposed a machine vision based inspection system for ball grid array defect inspection, which included joint position localization, region of interest (ROI) extraction, and range analysis model based failure analysis. They used many imaging processing methods such as Gaussian filter for noise reduction, Canny algorithm for edge contour detection and so on. However, the component is red ink dyed before capturing an image, which means this method is a destructive method. Gao \textit{et al.}~\cite{gao2016line}~proposed a defect detection method based on optical images for ball grid array by comparing features such as ball area, ball centroid bias, roundness, and a binary matrix indicating the existence of balls between the defect solder joint and the normal solder joint. With the help of the proposed line-based-clustering method, solder ball can be effectively segmented and recognized, which helps to extract the features mentioned above . The reference-free path-walking method proposed by Jin \textit{et al.}~\cite{jin2017reference}~also helped to extract the features for ball grid array defect detection.
    
    Automated X-ray inspection (AXI) uses X-rays as its source and the generated images can be used to inspect component solder joints whether they are defect or not. 
    Heavy materials that form the solder such as silver, copper, and lead are easier to be imaged and light materials that form other components as well as the board are not easily imaged under the X-ray. Thus, both solder external and internal characteristics can be imaged very well and structural defect can be inspected. Therefore, it has been widely used in electronics manufacturing to monitor the quality of printed circuit board. Yung \textit{et al.}~\cite{yung2018investigation}~proposed to use X-ray computerized tomography images for solder void defect detection by calculating solder void volume and void propagation direction with very high voxel size resolution from the X-ray system. It helped the process engineers understand the relationship between void defect from the X-ray imaging and the X-ray penetration energy. However, it had high complexity and took around one hour and ten minutes for one scanning sample, which made it difficult to implement within the manufacturing line. Jewler \textit{et al.}~\cite{jewler2019high}~used high resolution X-ray imaging for flip chip solder joint defect detection with high speed and sensitivity. However, the detection was based on visualizing the deviation of the defect joint features from the normal joint features in a 2D feature space, which was not an automatic detection method. 

    Recently, deep learning such as convolutional neural networks (CNNs) have shown outstanding performance on image based tasks such as image classification and object detection \cite{krizhevsky2012imagenet, zhang2019recent}. It automatically extracts features using local connectivity and weight sharing mechanism. Many methods have been proposed based on CNNs in solder joint defect detection. Wu \textit{et al.}~\cite{wu2019solder}~proposed to use CNN based object detection architecture, namely mask R-CNN method for joint defect detection. It could not only localize the joint, but also classify whether the solder joint was defect or not. Cai {et al.}~\cite{cai2018smt}~proposed to use three cascaded CNNs on optical images for solder joint defect detection. They incorporate the whole image as well as region of interest patches to classify whether a joint is qualified or unqualified. It achieved the best performance across traditional feature-extracted based models as well as feature-extraction free statistical models. More levels of models mean more manual labeling. It requires manual labeling not only for the whole image, but also for each image, four to five patches are required manual labeling. Goto \textit{et al.}~\cite{goto2019anomaly}~used adversarial convolutional autoencoder to extract features and exploited Hotelling’s T square to do anomaly detection. Compared with one-class support vector machine method, which is based on handmade features including substrate area, head in pillow area, circularity, and luminance ratio, the proposed method performs better in terms of false positive rate. Their method worked well on eight-slice samples with perfect isolated one solder joint on each image, which was not easy to obtain in the real world, where there could be varying number of slices and many solder joints could be captured by the X-ray detector in one image. 
    
    It is expected to save the workload of manual visual inspection using AXI. However, some built-in AXI algorithms nowadays are not efficient enough to correctly filtering out defect solder joints. 
Since few researches had merchandised AXI machine statistics, we calculated the statistics based a real-world dataset we obtained.
Among 518,292 through hole solder images that are labeled by an AXI machine as defects, only around 15\% are truly defects as labeled by specialist. The large number of false calls, the good solders that are incorrectly detected as defect ones, increases the specialist inspection workload. 
Therefore, there is a need to design an Artificial Intelligence (AI)-driven tool that can transfer the knowledge for specialist, automate the inspection, and hence reduce the workload of specialist. 

Current researches for X-ray imaging based solder defect detection rarely address the noises in the X-ray imaging dataset. The ROIs extracted by AXI may be incorrect and fail to enclose the solder joint area. Such noises can deteriorate solder defect detection. Furthermore, for different components, the solder joint need different depths of X-ray imaging for defect detection, which results in different number of X-ray imaging slices. 
Although there are some deep learning based methods implemented in PCB defect detection as mentioned, few researches focus on X-ray imaging and explore the methods for addressing the varying number of slices.

    In this paper, we mitigate the AXI machine inefficiency problem and decrease the specialist inspection workload afterwards by introducing AI to detect if a solder joint is defect or not. 
    Two AI based models are proposed, each of which consists of pre-processing method and deep learning model structure for detection. The pre-processing method is dedicatedly designed and two deep learning model structures are proposed. The proposed pre-processing method, namely channel-wise pre-processing method can address the varying number of slices problem and the incorrect ROI problem. The two model structures are designed based on 3D CNN and long short time memory (LSTM), which suits the proposed pre-processing methods. 
Since few deep learning based methods address the varying number of slices problem and the incorrect ROI problem in PCB X-ray imaging, the performance is compared between each other, and can be used as baseline for future researches.
The proposed methods have good generalization performance and its efficacy is verified through experimenting on a very large real-world 3D X-ray dataset. By incorporating the AI based models, $66.51\%$ normal joints are filtering out and only $33.49\%$ of normal joints are sent to the specialists for inspection, which reduces the specialist workload dramatically. The proposed methods are introduced in Section~\ref{sec:method}. The methods detail in Section~\ref{sec:pre} and Section~\ref{sec:structure}. 
Experiment and performance results are presented in Section~\ref{sec:exp} and we conclude the work in Section~\ref{sec:conclude}.

\section{Methodology}
\label{sec:method}
\begin{figure}[!htb]
    \centering
    \subfloat[Current inspection in production line]{
    	\label{fig:prod_line_a}
    	\includegraphics[width= 3.0in]{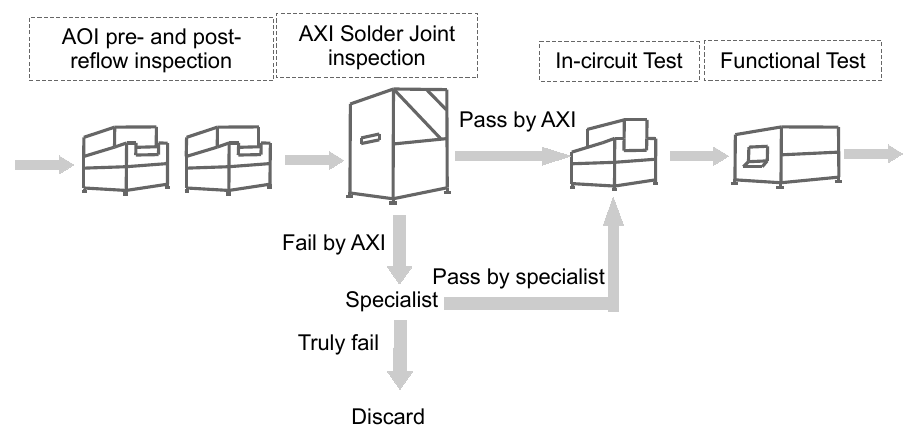}}\\
    \subfloat[Proposed AI-assisted inspection in production line]{
    	\label{fig:prod_line_b}
    	\includegraphics[width= 3.0in]{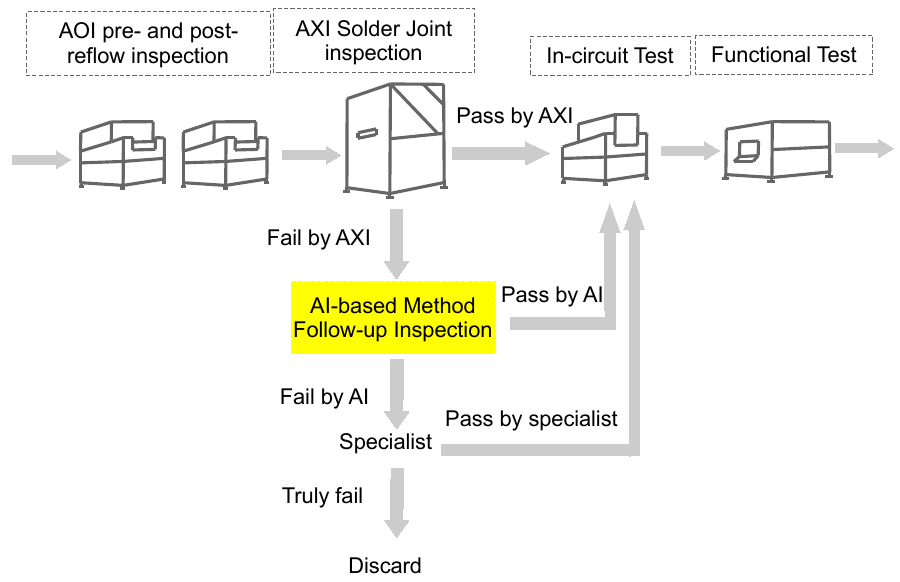}}
\caption{A production line with follow-up inspection after AXI}
\label{fig:production_line}
\end{figure}

In quality control of electronic production line, there are many tests as shown in Fig. \ref{fig:production_line}, including automated optical inspection (AOI), automated X-ray inspection (AXI), in-circuit test and functional test. Our work focuses on AXI, which detects defect solder joints based on X-ray imaging. Usually, solder joints that are detected as defect ones by AXI are sent to specialist for follow-up inspection as shown in Fig. \ref{fig:prod_line_a}. 
When AXI is inefficiency, specialists need to check many solder joints that are actually normal ones, which wastes a lot of time. Our work mitigates the inefficiency problem of AXI by introducing CNN based methods after AXI as shown in Fig. \ref{fig:prod_line_b}. Solder joints that are detected as defect by our methods will be sent to specialists, while normal joints are sent for next-step tests.

\subsection{3D Automatic X-ray Inspection}
\label{sub:3d}

Before introducing the proposed methods, it is obliged to introduce some background of the 3D automatic X-ray inspection (AXI) system we focus on, which helps understand the reasoning of the methods. 

\begin{figure}[!t]
\centering
\subfloat[Slice 00]{\includegraphics[width=1.2in]{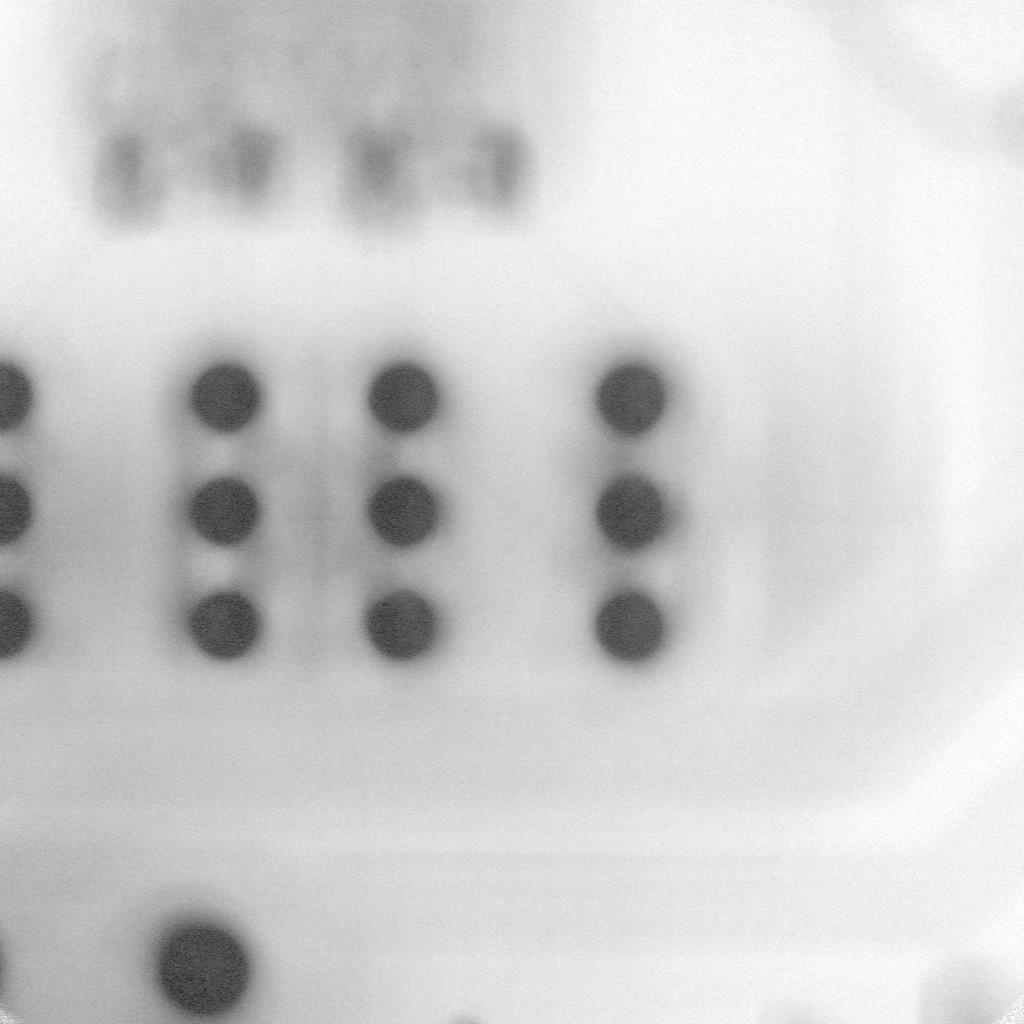}
\label{fig:sample_slice00}}
\hfil
\subfloat[Slice 01]{\includegraphics[width=1.2in]{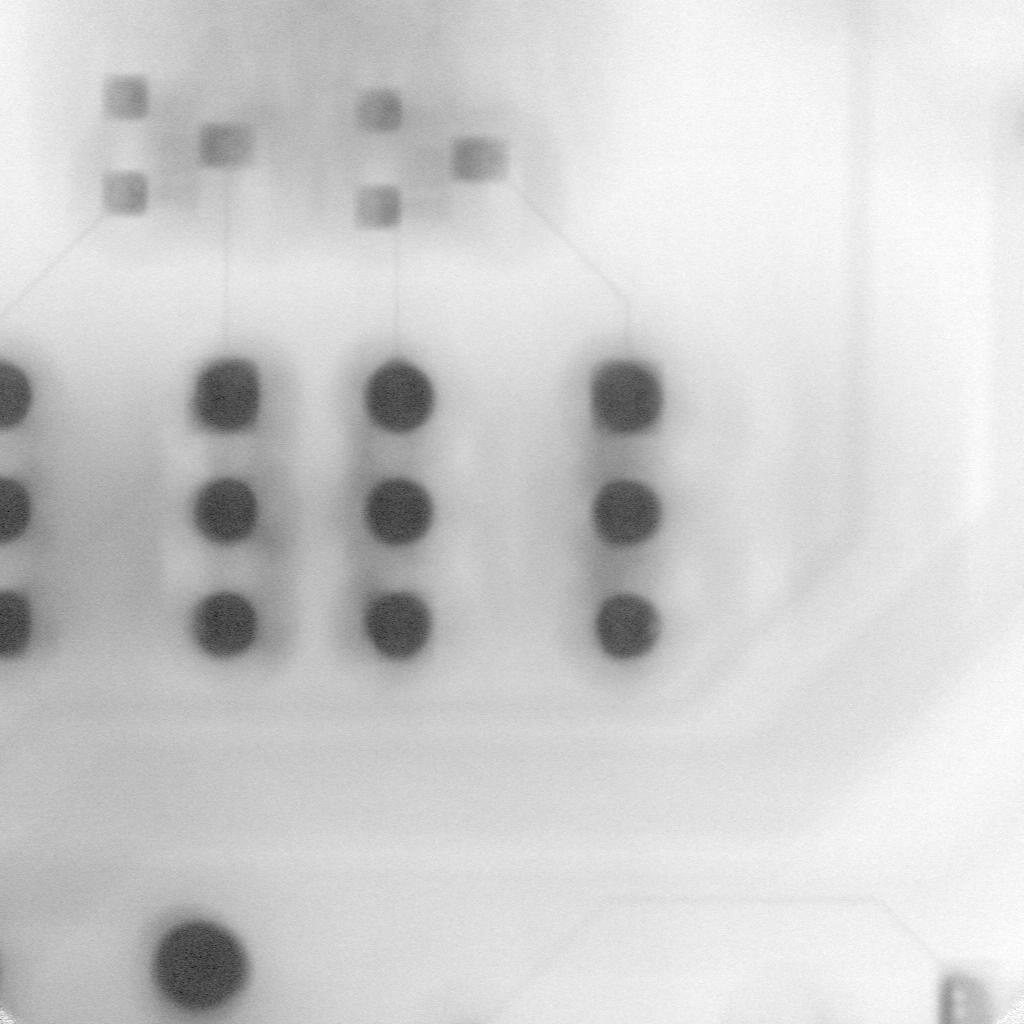}
\label{fig:sample_slice01}}
\vfil
\subfloat[Slice 02]{\includegraphics[width=1.2in]{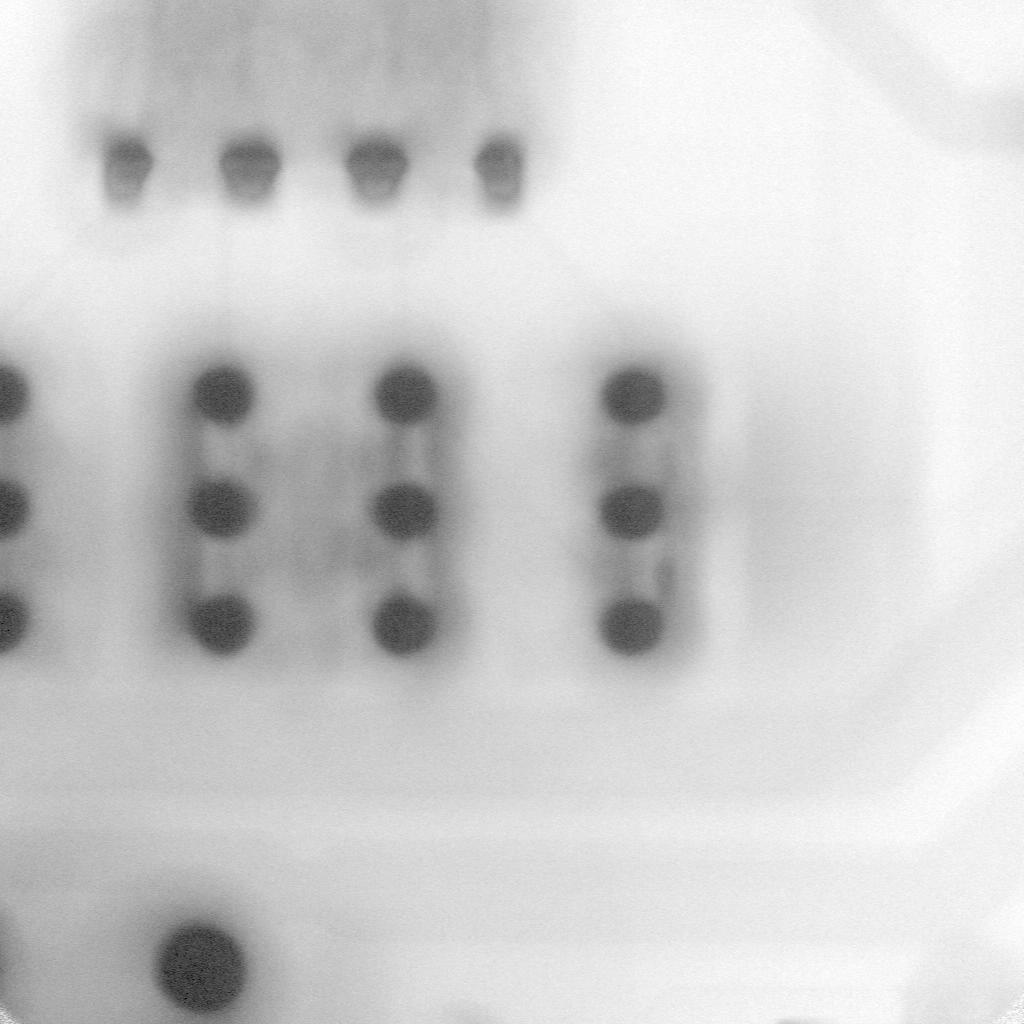}
\label{fig:sample_slice02}}
\hfil
\subfloat[Slice 03]{\includegraphics[width=1.2in]{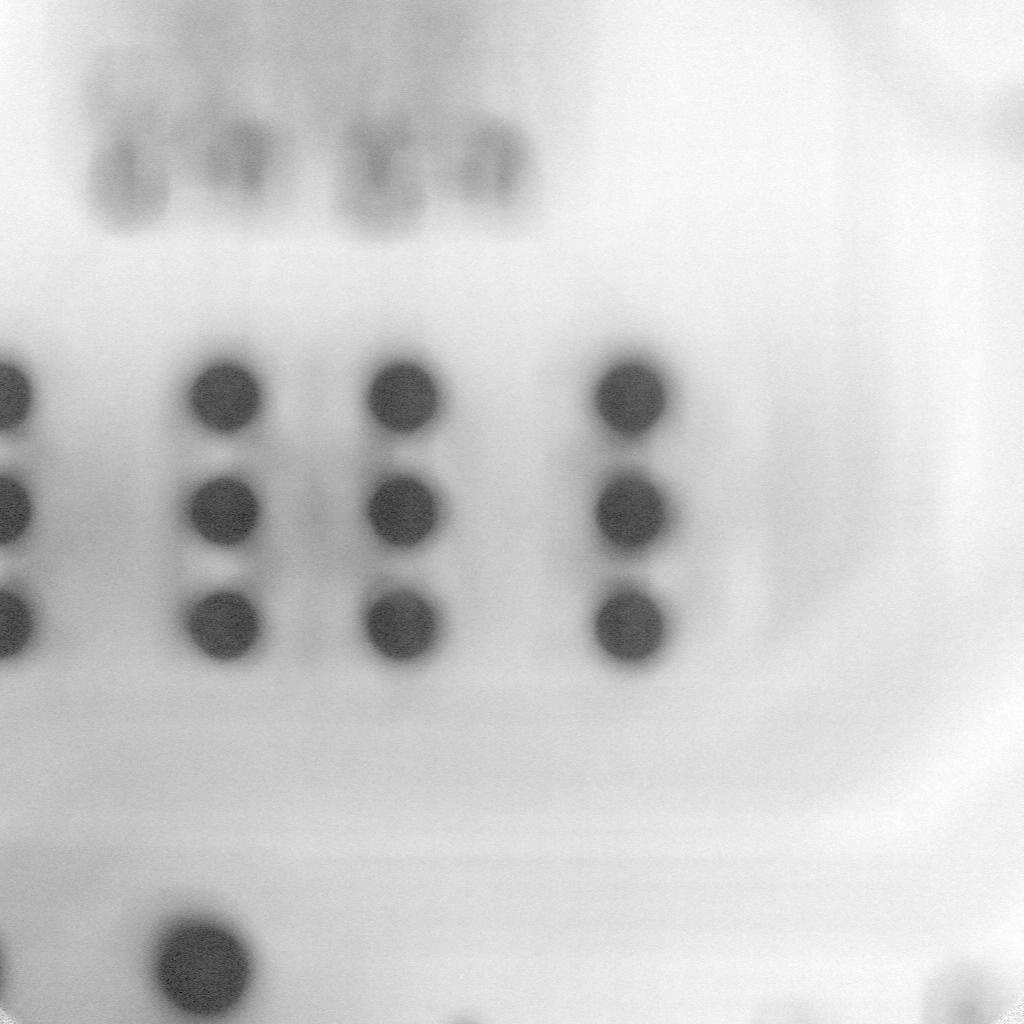}
\label{fig:sample_slice03}}
\caption{An example of 3D X-ray imaging with four slices.}
\label{fig:sample_slices}
\end{figure}

The 3D X-ray system can obtain different imaging depths of the PCB underneath the X-ray source with synchronized rotating source and the detector. The component joints on the focus plane would be sharp, while the ones above or below the focus plane would be blurred. Therefore, different slices regarding different depths of the PCB are obtained. These slices are important for joint defect detection as they show the hidden features of the joints. A example with four slices is shown in Fig. \ref{fig:sample_slices}. More slices provide more information for one solder joint. However, the label is given based on solder joint. That is to say, if one of the slices for the solder joint is defective, all the slices are labeled as defective. It is impossible to know which slices are defective. Therefore, all slices for one solder joint need to be treated together as one input. And different inputs have different number of slices.

\begin{figure}[!t]
\centering
\subfloat[]{\includegraphics[width=1.2in]{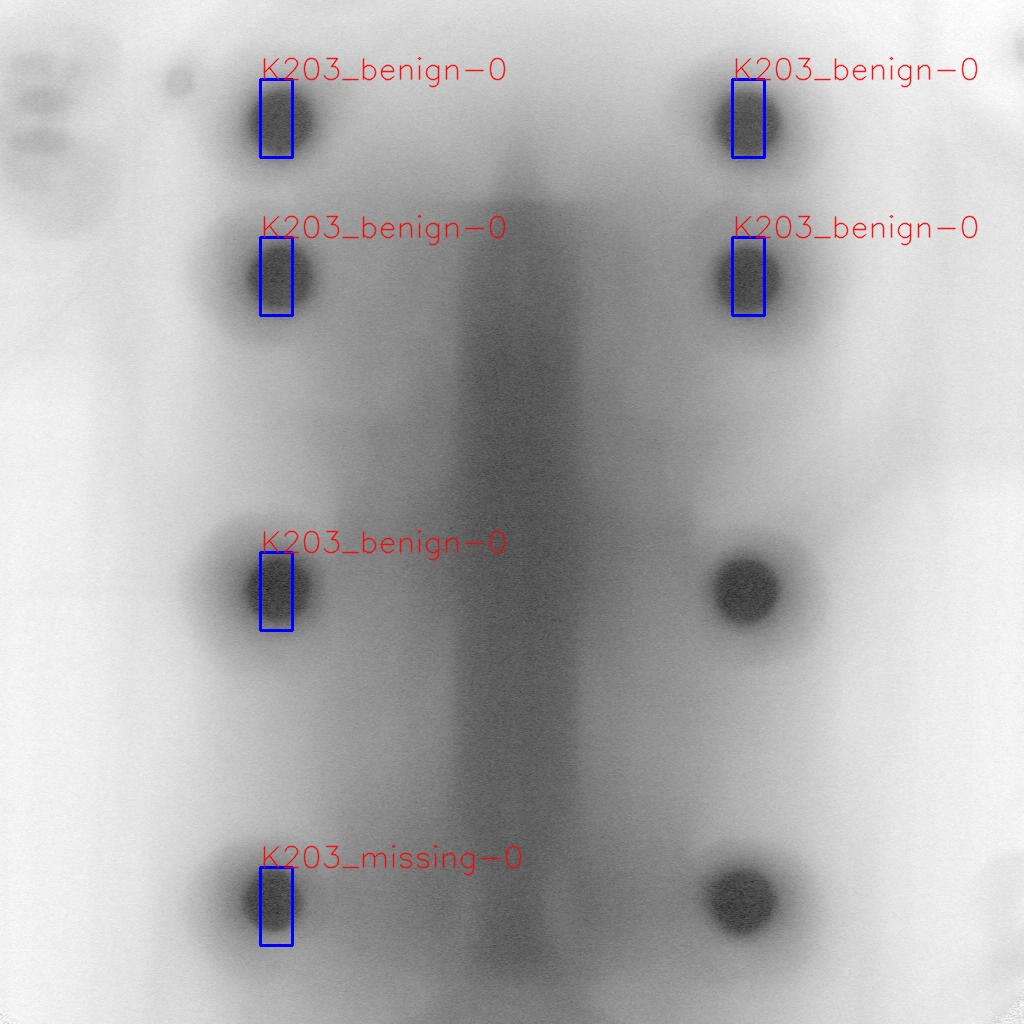}
\label{fig:sample_roi1}}
\hfil
\subfloat[]{\includegraphics[width=1.2in]{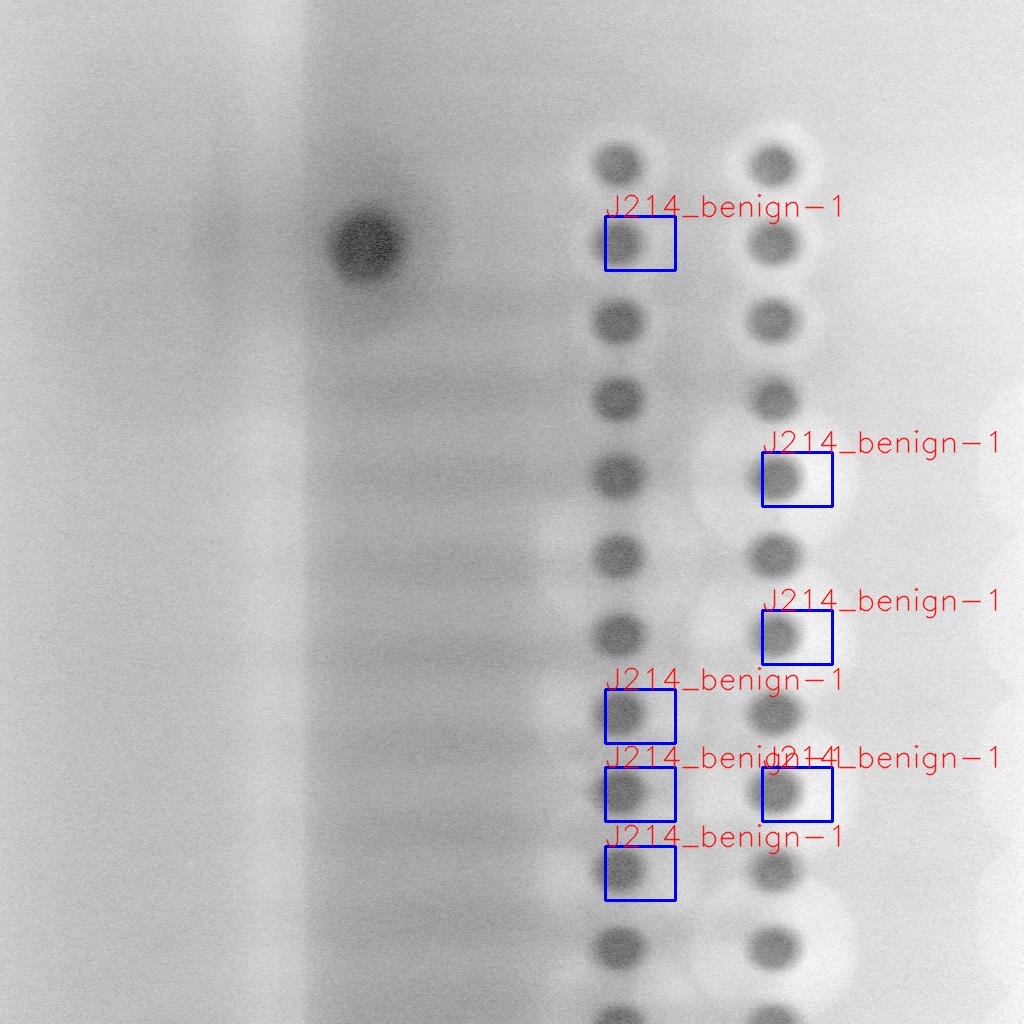}
\label{fig:sample_roi2}}
\caption{An example of AXI with ROIs.}
\label{fig:sample_rois}
\end{figure}

Along with the X-ray images, AXI also provides a bounding box of the inspected solder joint called region of interest (ROI). Ideally, ROIs not only provide the location of the inspected solder joint, but also the area of the inspected solder joint. However, some ROIs are not reliable. For example, ROIs fail to enclose the whole part of the solder joint due to the small size of ROI in Fig. \ref{fig:sample_roi1} and ROI shifts in Fig. \ref{fig:sample_roi2}. 

\subsection{Proposed Framework}
\begin{figure}[tb]
    \centering
    \includegraphics[width=2.5in]{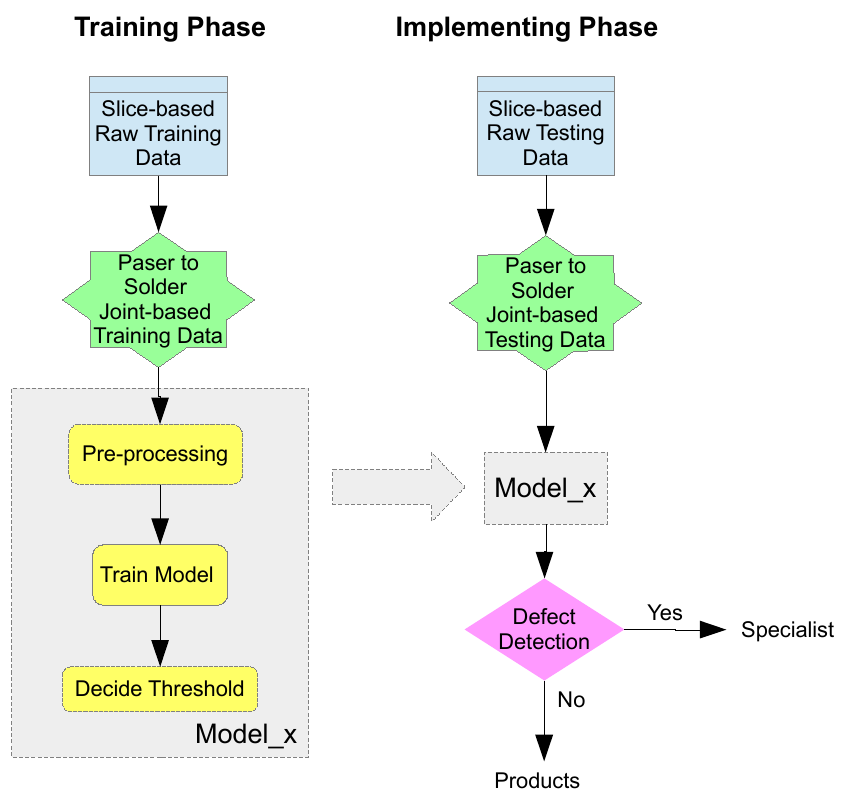}
    \caption{Proposed framework}
    \label{fig:framework}     
\end{figure}

The proposed defect solder detection framework consists with two phases, namely, the training phase and the implementing phase as shown in Fig. \ref{fig:framework}. 

During the training phase, raw mega data such as joint type, board type, ROI, and X-ray image file path from the machine is parsed first and prepared. Also, slice based image data is parsed into solder joint based image data. Then the image file is pre-processed before model training. In order to reduce false normal solders that should have been detected as defect solders, thresholding is empirically selected. 

During the implementing phase, the new incoming parsed and pre-processed data is sent to the trained model and compared with the threshold. Detected defects are passed to the operator for further analysis and decision making. 

Under this framework, two models are proposed. 
The proposed models have different AI model structures as shown in Table \ref{tab:methods}. Two AI model structures are proposed that fall into the categories of 3D CNN and LSTM.
In the following sections, the pre-processing method and two AI model structures are illustrated in detail.

\begin{table}[!htb]
 	\small
	\centering
	\captionsetup{labelsep=newline,singlelinecheck=false}
    	\begin{threeparttable}
	\caption{Model description}
	\label{tab:methods}
	\begin{tabular}{p{2.6cm}p{2.0cm}p{3.2cm}}
		\toprule
                Name & Pre-processing & Structure \\
                \hline
                3D CNN Model & Channel-wise & 3D CNN structure\\
                \hline
                LSTM based Model & Channel-wise & LSTM based structure\\
		\bottomrule
	\end{tabular}
	\end{threeparttable}
\end{table}

\section{Channel-wise Pre-processing Method}
\label{sec:pre}
Varying number of slices and noised ROIs are the two problems to handle. Based on the AXI characteristics mentioned in Section~\ref{sub:3d}, there are different number of slices for different solder joints, since some can be detected with deep depth of imaging, while some need both deep and shallow depths of imaging. The other problem is that some ROIs are not correct, which cannot surround the area of the joint of interest. We address the two problems by proposing the Channel-wise pre-processing method that concatenate all the slices in channel-wise direction before sent into the model.

Since there are varying number of slices, at first we pad zero-slices to the maximum number of slices, which is six in our case. 
As the dataset not only include the focused solders, but also the surrounding solders as well as the background, an ROI based cropping is implemented during the pre-processing step to reduce the size of input. 
The channel-wise pre-processing algorithm is shown in Algorithm~\ref{alg:6channel_pre}.

\begin{algorithm}
\LinesNumbered
	\KwIn{Original image dataset $\mathcal{D}_{image} = \{s_n^i; i~is~slice~number,~n~is~solder~number\}$, ROI = \{$xmin, xmax, ymin, ymax$\}}
	\KwOut{Solder joint patch dataset $\mathcal{D}_{patch}$, cropping ROI = \{$cxmin, cxmax, cymin, cymax$\}}
	$\mathcal{D}_{patch} \leftarrow \emptyset$\;
	\For{$k \leftarrow 1~to~n$}{
		\If{$s_k^i, i \in [1, 6]$ not exists}{
		Pad zero-slice to $s_k^i$\;
		}
		$c_k(x, y) = (\frac{xmax - xmin}{2} + xmin, \frac{ymax - ymin}{2}+ymin)$\;
		$cropLength = 1.5 \times \max{(xmax - xmin, ymax - ymin)}$\;
		$(cxmin, cymin) = center_k(x, y) - \frac{cropLength}{2}$\;
		$(cxmax, cymax) = center_k(x, y) + \frac{cropLength}{2}$\;
		\If{$(cxmin or cymin) \leq 0$ or $(cxmax or cymax)\geq 1024$}{
		pad zeros on $s_k$\;
		}
		$patch_k$ = Crop $s_k^{i}$ according to cropping ROI, $i\in[1,6]$\;
		Resize $patch_k$ to $128\times128$\;
		Add $patch_k$ to $\mathcal{D}_{patch}$\;
	}
\caption{Channel-wise pre-processing algorithm} 
\label{alg:6channel_pre}
\end{algorithm}

\begin{figure}[tb]
    \centering
    \includegraphics[width=2.2in]{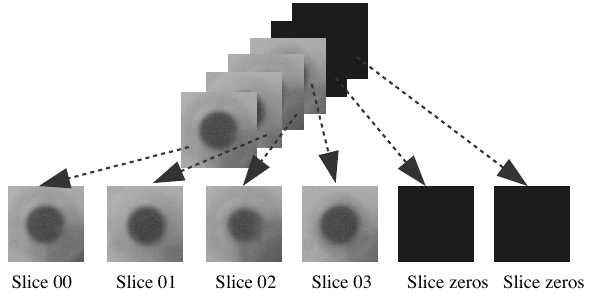}
    \caption{Channel-wise pre-processing sample}
    \label{fig:6channelpreprocess_sample}     
\end{figure}

The output of pre-processing consists of six channels as shown in Fig. \ref{fig:6channelpreprocess_sample}. Each channel represents one slice of the solder joint. For the slice amount that is less than six, zero-value slices are padded. 

\section{Deep Learning Model Structures}
\label{sec:structure}
Two deep learning model structures are proposed in this work. First one is designed based on 3D convolution and the second one is designed with LSTM module. 

\subsection{3D CNN Structure}
\label{sub:3d_cnn}
Compared with CNN, 3D CNN's kernel moves along three axes to extract features of the input. 
Since there are multiple slices for one solder joint, the solder joint slices are stacked along the third dimension. The 3D CNN architecture is detailed in Table \ref{tab:3d_cnn}.

\begin{table}[!htb]
    \small
	\centering
	\captionsetup{labelsep=newline,singlelinecheck=false}
    	\begin{threeparttable}
	\caption{3D CNN structure}
	\label{tab:3d_cnn}
	\begin{tabular}{p{2.8cm}p{2.5cm}p{2.5cm}}
		\toprule
		Layer name & Output shape & Kernel/weight size \\
		\hline
		Input 				& $128\times128\times6\times1$ 	& - \\
		\hline
		Conv3D				& $126\times126\times5\times8$ 	& $3\times3\times2\times8$  \\
		Conv3D				& $124\times124\times4\times16$ 	& $3\times3\times2\times16$  \\
		Max pooling			& $62\times62\times2\times16$		& -  \\
		\hline
		Conv3D				& $60\times60\times1\times32$ 	& $3\times3\times1\times32$  \\
		Conv3D				& $58\times58\times1\times64$ 	& $3\times3\times1\times64$  \\
		Max pooling			& $29\times29\times1\times64$		& -  \\
		Batch normalization		& $29\times29\times1\times64$		& -  \\
		\hline
		Dropout				& $53824$					& -\\
		Fully connected			& $1024$						& $53824\times1024$ \\
		Dropout				& $1024$						& -\\
		Fully connected 		& $2$						& $1024\times2$\\
		\bottomrule
	\end{tabular}
	\end{threeparttable}
\end{table}

\subsection{LSTM Based Structure}
LSTM neural network is one of the recurrent neural networks that can model a sequence of inputs. By inserting the LSTM module with an input gate, an output gate, and a forget gate, the neural network is able to capture the relationship within the sequential inputs. 
For our case, slices from 00 to 05 are the sequential inputs. Since there are six slices, the LSTM module continuously computing six times until all the slices are processed.

\begin{figure}[tb]
    \centering
    \includegraphics[width=3.2in]{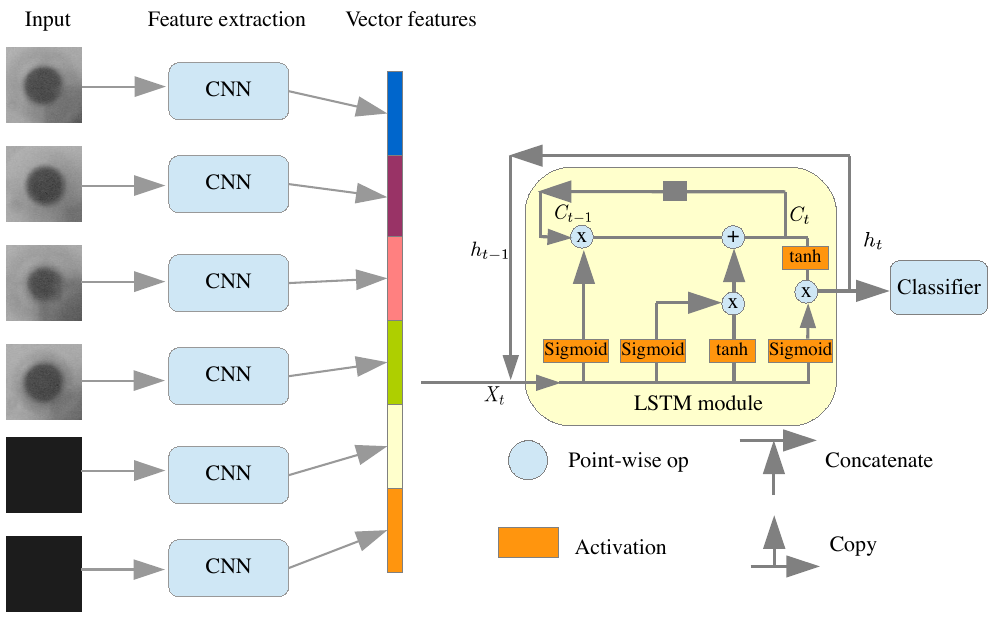}
    \caption{LSTM based structure illustration}
    \label{fig:lstm_based}     
\end{figure}

Since images are high dimension inputs, features are extracted by CNNs from each slice before sent into the LSTM module as shown in Fig. \ref{fig:lstm_based}. The features extracted from the CNNs are in vector forms and each vector consists of the information from each slice. The LSTM module is followed a classifier composed with two fully connected layers, which are used for classification as shown in Table \ref{tab:lstm}. 

\begin{table}[!htb]
    \small
	\centering
	\captionsetup{labelsep=newline,singlelinecheck=false}
    	\begin{threeparttable}
	\caption{LSTM based structure}
	\label{tab:lstm}
	\begin{tabular}{p{2.5cm}p{2.0cm}p{2.5cm}}
		\toprule
		Layer name & Output shape & Kernel/weight size \\
		\hline
		LSTM 				& $2048$ 						& - \\
		\hline
		Fully connected			& $512$						& $2048\times512$ \\
		Dropout				& $512$						& -\\
		Fully connected 		& $2$						& $512\times2$\\
		\bottomrule
	\end{tabular}
	\end{threeparttable}
\end{table}

\section{Experiment and results}
\label{sec:exp}
We implement the two models using Python Keras and TensorFlow. 
Adam is used as our optimization method. 
The activation function is the rectified linear function (ReLU) \cite{krizhevsky2012imagenet}. The learning rate for 3D CNN and LSTM based model in Table \ref{tab:methods} is set $1e-5$ with decay $1e-6$.
\subsection{Dataset and Metrics}
The dataset slice distribution and normal-defect ratio show in Table \ref{tab:dataset}. 
Training dataset, validation dataset and testing dataset are divided based on board types for testing generalization performance. That is to say the board types in the testing phase are not seen during training the models to mimic the real-world case. 
    In the real-world scenario, benign solder joints are far more than defect solder joints. To deal with data imbalanced problem, we down sampling the benign data to get a balanced training dataset. 
\begin{table}[!htb]
 	\small
	\centering
	\captionsetup{labelsep=newline,singlelinecheck=false}
    	\begin{threeparttable}
	\caption{Dataset slice distribution}
	\label{tab:dataset}
	\begin{tabular}{p{2.5cm}p{2.5cm}}
		\toprule
		\multicolumn{2}{c}{Pin Through Hole Solder}\\
		\hline
    			\multirow{6}{*}{Slice distribution} 
				   & 1 slice: 763 \\
    				~ & 2 slices: 1694 \\
				~ & 3 slices: 7576 \\
				~ & 4 slices: 466029 \\
				~ & 5 slices: 41345 \\
				~ & 6 slices: 945 \\
              	\hline
                Normal-defect ratio & 1:0.18 \\
		\bottomrule
	\end{tabular}
	\end{threeparttable}
\end{table}
    
Accuracy is one of the commonly used metrics. However, under the condition of imbalanced problem, as the test data includes large amount of benign data and small amount of anomaly data, accuracy is not fair for performance checking. 
For example, one method can classify all the solder joints as benign, which will give a very high accuracy rate. But it can not detect any defect solder joint. Instead, $Recall$, false positive rate~($FPR$), and area under the receiver operating characteristic~($AUROC$) are commonly used.
 
The true positive rate ($TPR$) is defined as correct positive results that happen in all positive samples, which is also called $Recall$. $FPR$ is defined as incorrect positive results that happen in all negative samples. In our case, positive is defined as defect solder joint and negative is defined as normal solder joint.

The receiver operating characteristic (ROC) curve is a commonly used tool to analyze and visualize the performance of a binary classifier as its discrimination threshold is varied. 
ROC curve takes $FPR$ as its horizontal axis and takes $TPR$ as its vertical axis. 
Classifier with upper left conner points has better overall performance. The area under the $ROC$ is $AUROC$ that measures the general performance regardless of the thresholds. 

Compared with metric of accuracy, which measures the proportion between the sum of $TP$ and $TN$ and the total number of instances, $Recall$, $FPR$, and $AUROC$ are more precise as they take imbalance classes of data into consideration.

\subsection{Results}
\label{sec:result}
Results of two models show in Table \ref{tab:3dcnn_model} and Table \ref{tab:lstm_model}. From the tables, there is a tradeoff between $Recall$ and $FPR$. As threshold increases, the model can achieve higher $Recall$, and $FPR$ increases respectively. For different models, the threshold has different impacts. For example, for 3D CNN model, when threshold is $0.1$, on validation dataset, its $Recall = 0.9795$, while for LSTM based model, $Recall = 0.9826$. 

As the model generalizes from validation dataset to the testing dataset, there is around $2\%\sim5\%$ drop for $Recall$ and $1\%\sim8\%$ increase for $FPR$. The empirical threshold can be picked based on the validation dataset. In Table \ref{tab:t_comparison}, in order to achieve validation dataset $Recall = 0.90$, thresholds for 3D CNN model and LSTM based model are $0.34$ and $0.28$ respectively. LSTM based model has higher $Recall = 0.8836$ for testing dataset. Compared with $Recall = 0.90$ for validation dataset, there is a drop of around $2\%$ when generalizing from validation dataset to testing dataset. For 3D CNN model, the drop is around $5\%$. 

\begin{table}[!htb]
	 \small
	\centering
	\captionsetup{labelsep=newline,singlelinecheck=false}
    	\begin{threeparttable}
	\caption{3D CNN performance}
	\label{tab:3dcnn_model}
	\begin{tabular}{p{1.0cm}p{1.0cm}p{1.0cm}p{1.0cm}p{1.0cm}}
		\toprule
                \multirow{2}{*}{Threshold} &
                  \multicolumn{2}{c}{Validation} &
                  \multicolumn{2}{c}{Testing} \\
                & Recall & FPR & Recall & FPR \\
                \hline
                0.1 & 0.9795 & 0.4204 & 0.9200 & 0.4240 \\
                \hline
                0.2 & 0.9457 & 0.3018 & 0.8884 & 0.2798 \\
                \hline
                0.3 & 0.9149 & 0.2248 & 0.8654 & 0.2058 \\
                \hline
                0.4 & 0.8797 & 0.1729 & 0.8402 & 0.1582  \\
                \hline
                0.5 & 0.8445 & 0.1365 & 0.8146 & 0.1255 \\
		\bottomrule
	\end{tabular}
	\end{threeparttable}
\end{table}
\begin{table}[!htb]
	 \small
	\centering
	\captionsetup{labelsep=newline,singlelinecheck=false}
    	\begin{threeparttable}
	\caption{LSTM based model performance}
	\label{tab:lstm_model}
	\begin{tabular}{p{1.0cm}p{1.0cm}p{1.0cm}p{1.0cm}p{1.0cm}}
		\toprule
                \multirow{2}{*}{Threshold} &
                  \multicolumn{2}{c}{Validation} &
                  \multicolumn{2}{c}{Testing} \\
                & Recall & FPR & Recall & FPR \\
                \hline
                0.1 & 0.9826 & 0.4745 & 0.9556 & 0.5568 \\
                \hline
                0.2 & 0.9535 & 0.3597 & 0.9134 & 0.3574 \\
                \hline
                0.3 & 0.9213 & 0.2564 & 0.8754 & 0.2433 \\
                \hline
                0.4 & 0.8830 & 0.1849 & 0.8508 & 0.1800  \\
                \hline
                0.5 & 0.8480 & 0.1378 & 0.8261 & 0.1362 \\
		\bottomrule
	\end{tabular}
	\end{threeparttable}
\end{table}


\begin{table}[!htb]
 	\small
	\centering
	\captionsetup{labelsep=newline,singlelinecheck=false}
    	\begin{threeparttable}
	\caption{Threshold comparison}
	\label{tab:t_comparison}
	\begin{tabular}{p{2.0cm}p{1.5cm}p{1.0cm}p{1.0cm}}
		\toprule
                Threshold (val) & Model & Recall & FPR \\
                \hline
                0.34 & 3D CNN & 0.8556 & 0.1849  \\
                \hline
                0.28 & LSTM & 0.8836 & 0.2908  \\
		\bottomrule
	\end{tabular}
	\end{threeparttable}
\end{table}

\begin{figure}[tb]
    \centering
    \includegraphics[width=3in]{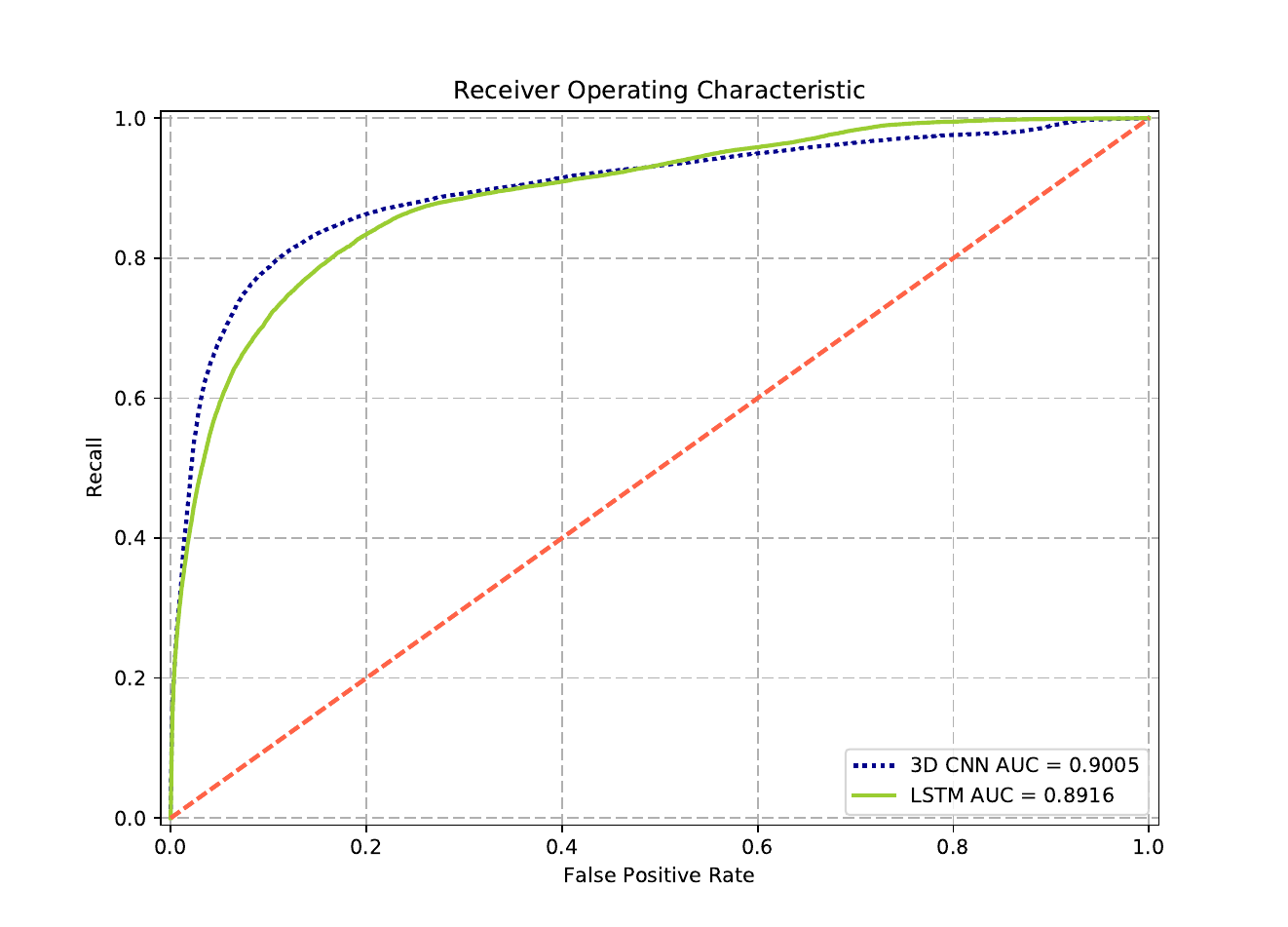}
    \caption{Overall performance of 3D CNN and LSTM based models}
    \label{fig:roimask_roc}     
\end{figure}

The overall performance of the proposed models can be seen in Fig. \ref{fig:roimask_roc}. Regardless of thresholds, two models perform similarly in terms of $AUROC$. 
In reality, some industrial requirement is expected such as $Recall \geqslant 0.90$ and $Recall \geqslant 0.95$. 
For less strict requirement ($Recall \geqslant 0.90$), 3D CNN achieves the lower $FPR = 0.3349$ as shown in Table \ref{tab:comparison}. 
That is to say, in the production line, compared with sending all the normal solder joints that detected as defects by the AXI machine to the specialist for manual inspection, only $33.49\%$ of normal joints are sent to the specialists and $66.51\%$ of normal joints are filtering out, which reduces the specialist workload dramatically. 
For strict requirement ($Recall \geqslant 0.95$), LSTM based model has the lower $FPR = 0.5872$ that almost reduces the specialist workload by $50\%$.

\begin{table}[!htb]
 	\small
	\centering
	\captionsetup{labelsep=newline,singlelinecheck=false}
    	\begin{threeparttable}
	\caption{Performance comparison}
	\label{tab:comparison}
	\begin{tabular}{p{1.5cm}p{1.0cm}p{2.3cm}p{2.3cm}}
		\toprule
                Model & AUROC & FPR@$90\%$Recall & FPR@$95\%$Recall\\
                \hline
                3D CNN & 0.9005 & 0.3349 & 0.6040\\
                \hline
                LSTM & 0.8916 & 0.3637 & 0.5872\\
		\bottomrule
	\end{tabular}
	\end{threeparttable}
\end{table}

\section{Conclusion}
\label{sec:conclude}
In this work, two models are proposed for follow-up defect inspection. Channel-wise pre-processing method is proposed to address the varying number of slices problem and the incorrect ROI problem, and two deep learning model structures are proposed that suit the pre-processing method.
The proposed two models have similar general performances in terms of $AUROC$. For less strict $Recall$ requirement, 3D CNN performs better, while for strict $Recall$ requirement, LSTM based model performs better. 
By introducing AI into the production line, specialist workload can be dramatically reduced. The performance can be the baseline for future researches on the X-ray imaging defect detection problems.


%

%

\section*{Acknowledgment}
This research work was in part supported by the China Scholarship Council, Keysight Technologies, and NSFC Projects of International
Cooperation and Exchanges (Project No. 61850410535).

\ifCLASSOPTIONcaptionsoff
  \newpage
\fi



\bibliographystyle{IEEEtran} 
\bibliography{jointAD} 

\begin{thebibliography}{10}
\providecommand{\url}[1]{#1}
\csname url@samestyle\endcsname
\providecommand{\newblock}{\relax}
\providecommand{\bibinfo}[2]{#2}
\providecommand{\BIBentrySTDinterwordspacing}{\spaceskip=0pt\relax}
\providecommand{\BIBentryALTinterwordstretchfactor}{4}
\providecommand{\BIBentryALTinterwordspacing}{\spaceskip=\fontdimen2\font plus
\BIBentryALTinterwordstretchfactor\fontdimen3\font minus
  \fontdimen4\font\relax}
\providecommand{\BIBforeignlanguage}[2]{{%
\expandafter\ifx\csname l@#1\endcsname\relax
\typeout{** WARNING: IEEEtran.bst: No hyphenation pattern has been}%
\typeout{** loaded for the language `#1'. Using the pattern for}%
\typeout{** the default language instead.}%
\else
\language=\csname l@#1\endcsname
\fi
#2}}
\providecommand{\BIBdecl}{\relax}
\BIBdecl

\bibitem{leinbach2001and}
G.~Leinbach and S.~Oresjo, ``The why, where, what, how, and when of automated
  x-ray inspection,'' \emph{Agilent Technologies}, 2001.

\bibitem{lu2018detection}
X.~Lu, Z.~He, L.~Su, M.~Fan, F.~Liu, G.~Liao, and T.~Shi, ``Detection of micro
  solder balls using active thermography technology and k-means algorithm,''
  \emph{IEEE Transactions on Industrial Informatics}, vol.~14, no.~12, pp.
  5620--5628, 2018.

\bibitem{huang2019developing}
C.-Y. Huang, J.-H. Hong, and E.~Huang, ``Developing a machine vision inspection
  system for electronics failure analysis,'' \emph{IEEE Transactions on
  Components, Packaging and Manufacturing Technology}, vol.~9, no.~9, pp.
  1912--1925, 2019.

\bibitem{gao2016line}
H.~Gao, W.~Jin, X.~Yang, and O.~Kaynak, ``A line-based-clustering approach for
  ball grid array component inspection in surface-mount technology,''
  \emph{IEEE Transactions on Industrial Electronics}, vol.~64, no.~4, pp.
  3030--3038, 2016.

\bibitem{jin2017reference}
W.~Jin, W.~Lin, X.~Yang, and H.~Gao, ``Reference-free path-walking method for
  ball grid array inspection in surface mounting machines,'' \emph{IEEE
  Transactions on Industrial Electronics}, vol.~64, no.~8, pp. 6310--6318,
  2017.

\bibitem{yung2018investigation}
L.~C. Yung, ``Investigation of the solder void defect in ic semiconductor
  packaging by 3d computed tomography analysis,'' in \emph{2018 IEEE 20th
  Electronics Packaging Technology Conference (EPTC)}.\hskip 1em plus 0.5em
  minus 0.4em\relax IEEE, 2018, pp. 886--889.

\bibitem{jewler2019high}
S.~J. Jewler, ``High resolution automatic x-ray inspection for continuous
  monitoring of advanced package assembly,'' in \emph{2019 International Wafer
  Level Packaging Conference (IWLPC)}.\hskip 1em plus 0.5em minus 0.4em\relax
  IEEE, 2019, pp. 1--5.

\bibitem{krizhevsky2012imagenet}
A.~Krizhevsky, I.~Sutskever, and G.~E. Hinton, ``Imagenet classification with
  deep convolutional neural networks,'' in \emph{Advances in neural information
  processing systems}, 2012, pp. 1097--1105.

\bibitem{zhang2019recent}
Q.~Zhang, M.~Zhang, T.~Chen, Z.~Sun, Y.~Ma, and B.~Yu, ``Recent advances in
  convolutional neural network acceleration,'' \emph{Neurocomputing}, vol. 323,
  pp. 37--51, 2019.

\bibitem{wu2019solder}
H.~Wu, W.~Gao, and X.~Xu, ``Solder joint recognition using mask r-cnn method,''
  \emph{IEEE Transactions on Components, Packaging and Manufacturing
  Technology}, 2019.

\bibitem{cai2018smt}
N.~Cai, G.~Cen, J.~Wu, F.~Li, H.~Wang, and X.~Chen, ``Smt solder joint
  inspection via a novel cascaded convolutional neural network,'' \emph{IEEE
  Transactions on Components, Packaging and Manufacturing Technology}, vol.~8,
  no.~4, pp. 670--677, 2018.

\bibitem{goto2019anomaly}
K.~Goto, K.~Kato, S.~Nakatsuka, T.~Saito, and H.~Aizawa, ``Anomaly detection of
  solder joint on print circuit board by using adversarial autoencoder,'' in
  \emph{Fourteenth International Conference on Quality Control by Artificial
  Vision}, vol. 11172.\hskip 1em plus 0.5em minus 0.4em\relax International
  Society for Optics and Photonics, 2019, p. 111720T.

\end{thebibliography}
\end{document}